\documentclass[12pt]{iopart}

\usepackage{iopams}

\usepackage{graphicx}
\usepackage{amssymb}
\usepackage{epsfig}

\def\be{\begin{equation}}
\def\ee{\end{equation}}

\begin{document}

\title{Canonical Statistical Model and hadron production in $e^+e^-$ annihilations}

\author{K. Redlich$^{a,e}$, A. Andronic$^b$,   F.~Beutler$^c$  \\P. Braun-Munzinger$^{b,d,e}$ and J. Stachel$^c$}

\address{
$^a$ Institute of Theoretical Physics, University of Wroclaw, PL--50204 Wroc\l aw, Poland
\\
$^b$ GSI Helmholtzzentrum f\"ur Schwerionenforschung, D-64291
Darmstadt, Germany
\\
$^c$ Physikalisches Institut der Universit\"at Heidelberg, D-69120
Heidelberg, Germany
\\
$^d$ ExtreMe Matter Institute EMMI,  D-64291 Darmstadt, Germany
\\
$^e$ Technical University Darmstadt, D-64289 Darmstadt, Germany}

\begin{abstract}
We discuss the production of hadrons in $e^+e^-$ collisions at
$\sqrt s=91$ GeV. We address the question wether the particle
yields measured in the final states are consistent with the
statistical model predictions. In the model formulation we account
for exact conservation of all relevant quantum numbers
 using the canonical
description  of the  partition function. Within our model    the
validity of the thermodynamical approach to quantify particle
production in $e^+e^-$ annihilations is not obvious.
\end{abstract}

\vspace{2mm}

\section{Introduction}

One of the essential results in heavy ion collisions was the
observation that particle yields measured in a final state closely
resemble a thermal equilibrium population
\cite{aat,review,wetterich}.  The natural question was whether
this statistical behavior is a unique feature of high energy
nucleus-nucleus collisions or whether it is also applicable in
elementary collisions like, e.g., $e^{+}e^{-}$. Previous
publications \cite{Becattini:1995if} indicated that indeed hadron
production in $e^{+}e^{-}$ collisions can be well described within
a thermal model provided that local quantum number conservation is
properly implemented. In  view of the most complete  and extended
data summarized e.g. by the Particle Data Group (PDG) \cite{pdg}
the above question has been recently addressed independently in
\cite{our} and \cite{becattini08}.

In this contribution we discuss    hadron production in
$e^{+}e^{-}$ annihilations at $\sqrt s=91 $ GeV based on the
thermal model analysis of  \cite{our}. We also focus on different
implementations of the statistical model and discuss  the
importance of quantum statistic effects and the mass--cut in the
hadron mass spectrum. We quantify  the production of heavy flavors
and compare the model predictions with available data.

\section{The statistical model and charge conservation}

The usual form of the statistical model in the grand canonical
ensemble formalism cannot be used when the number of produced
charged particles is small. This is the case  if   either the
temperature $T$ or the volume $V$ or both are small. As a rule of
thumb one needs $VT^3>>1$ for a grand canonical description to be
applicable \cite{hagedorn}. In $e^+e^-$ annihilations where
modelling the particle production within a thermal approach
requires an exact formulation of conservation laws. In such a
system one needs to account for an exact conservation of five
quantum numbers: the baryon number $N$, strangeness $S$, electric
charge $Q$, charm $C$ and bottom $B$.

\begin{figure*}
\begin{center}
\includegraphics[width=6.7cm]{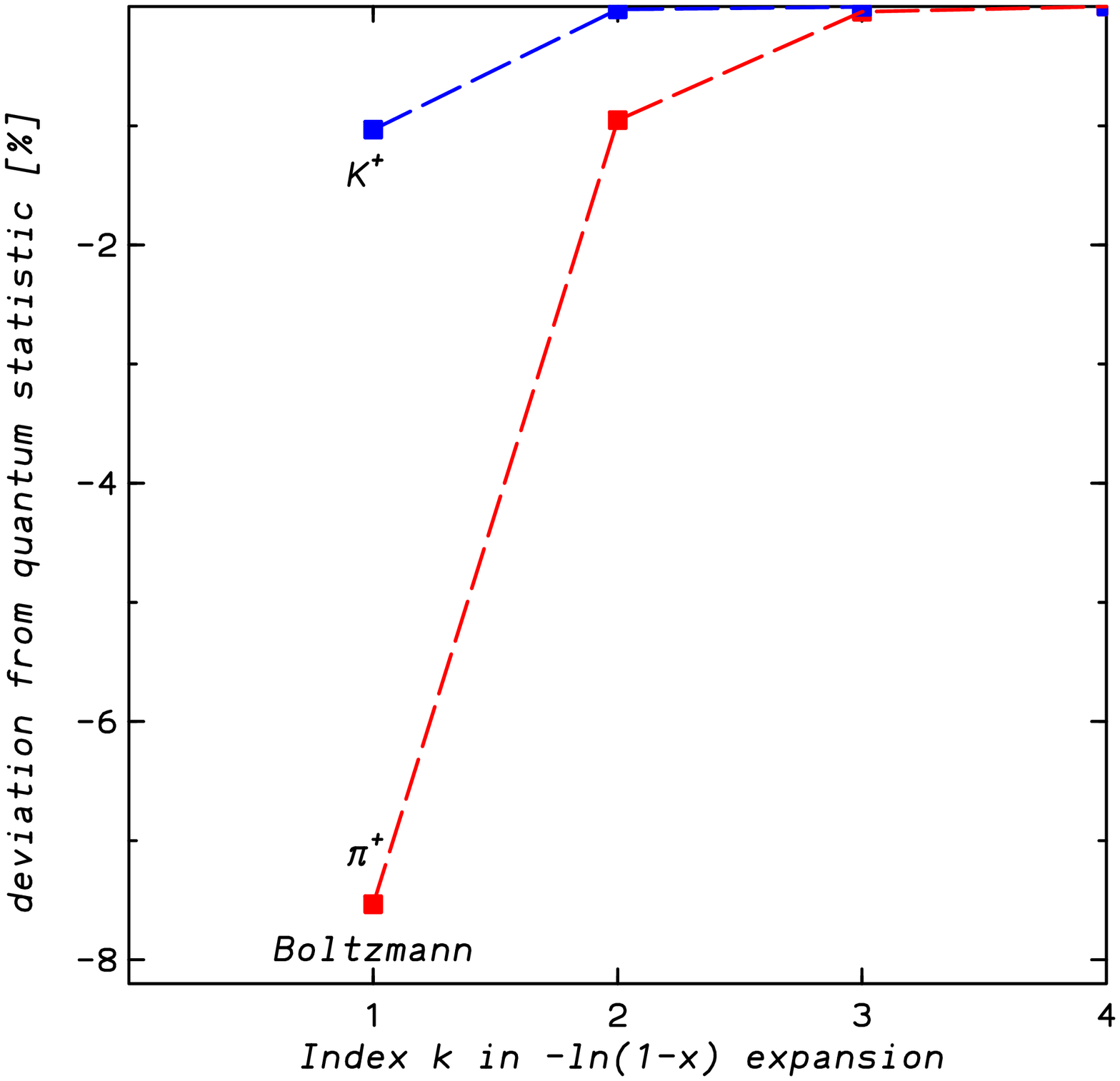}
\includegraphics[width=6.79cm]{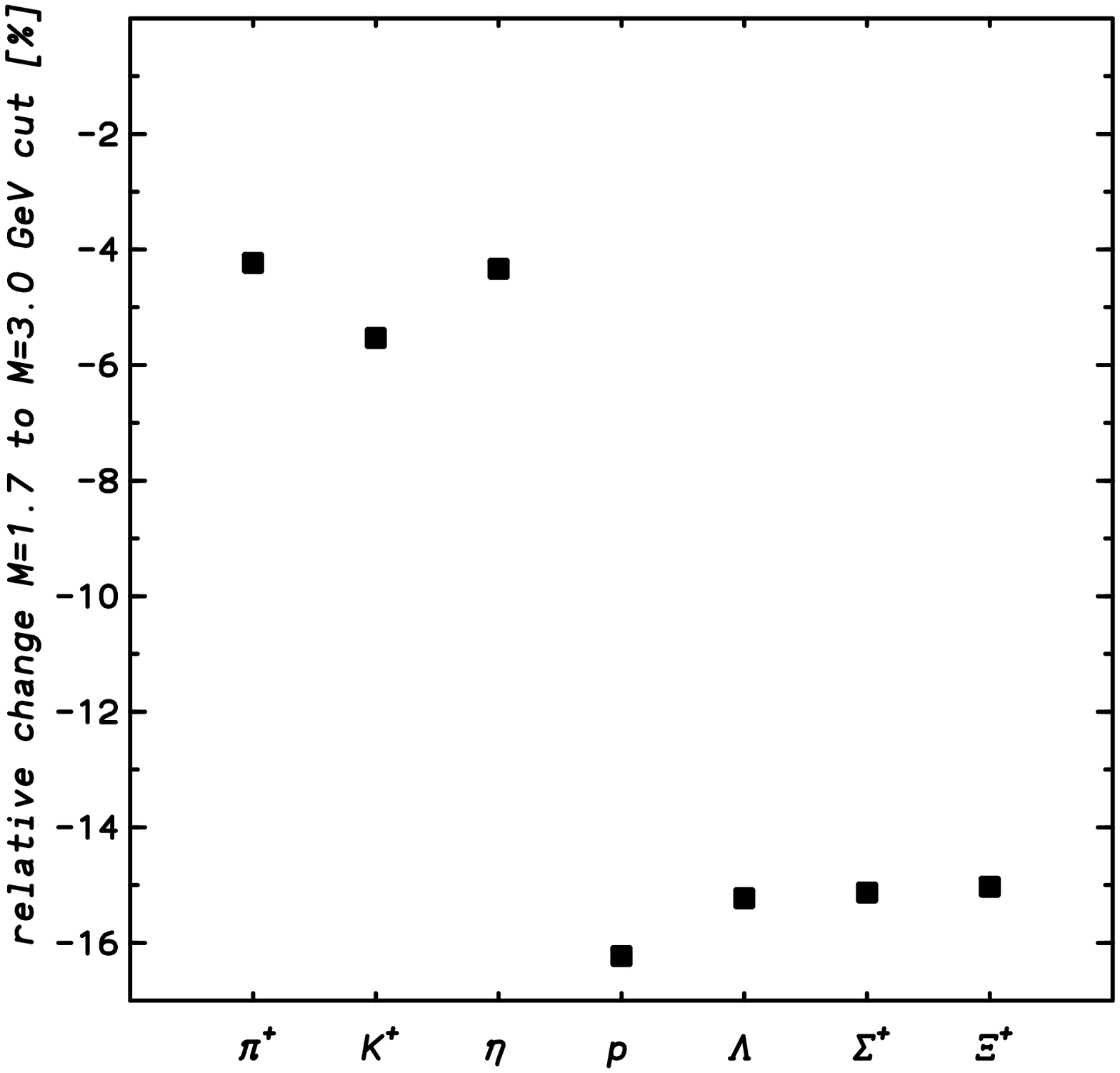} \caption{ The
(left--hand figure) shows  deviations of pion and kaon yields from
their exact quantum statistics values (see text). The $k=1$ term
corresponds to the Boltzmann approximation.  The (right--hand
figure): the relative change of  the particle yields calculated in
the hadron resonance gas model with a mass spectrum cut   at the
mass $M=1.7$ and $M=3.0$ GeV. }\label{fig1}
\end{center}
\end{figure*}

The appropriate tool to deal in a statistical mechanics framework
with a
 system  of quantum numbers  $\vec X=(N,S,Q,C,B)$
 is the canonical partition function \cite{hagedorn,turko}
\begin{equation}
Z_{N,S,Q,C,B} = \frac {1}{(2\pi)^5}\int\limits^{\pi}_{-\pi}
d^5\vec\phi\;e^{i\vec\phi\vec X} \exp{(\sum_j z_j)},
\label{eq:partition1}
\end{equation}
where
\begin{equation}
z_j= g_j\frac{V}{(2\pi)^3}\int d^3p\;\ln(1\pm
\exp{(-\sqrt{p^2+m_j^2}/T -i\vec x_j\vec \phi)})^{\pm 1},
\label{eq:partition1b}
\end{equation}
and  $x_j$ is a five  component vector $x_j=(N_j,S_j,Q_j,C_j,B_j)$
containing the quantum numbers of the  particle species $j$. The
quantity $\phi = (\phi_N,\phi_S,\phi_Q,\phi_C,\phi_B)$ is an
element  of the symmetry group $[U(1)]^5$ related with additive
conservation laws.  In this expression, each $\phi_X$ corresponds
to the conservation of the corresponding quantum number $ X$ and
$z_j$ is the single particle partition function for particle with
mass $m_j$, spin-isospin degeneracy factor $g_j$, and a system
with volume V and temperature T. The sum in
Eq.~(\ref{eq:partition1}) runs over all particle species in the
hadronic gas.

The integral representation of the partition function in
Eq.~(\ref{eq:partition1}) is not convenient for numerical analysis
as the integrand is a strongly oscillating function. Thus, we
first  expand   the logarithm

\begin{equation}
\ln(1\pm x)^{\pm 1}=\sum_{k=1}^{\infty}(\pm1)^{k+1}{{x^k}\over k}
\label{eq:log}
\end{equation}
and then, using the method described in
\cite{Cleymans:1997ib,pbm4,beutler_diploma}, we express the
partition function ~ (\ref{eq:partition1}) in a  series of Bessel
functions  to obtain a result that is free from oscillations.
Furthermore, from Eq.~(\ref{eq:partition1}) we obtain the
multiplicity $ \langle n_j\rangle$  for particle
  species j by introducing a
fugacity parameter $\lambda_j$ which multiplies the particle
partition function $z_j$ and by differentiating
\begin{equation}
\left.\langle n_j\rangle = \frac{\partial\ln Z}{\partial
\lambda_j}\right|_{\lambda_j=1}. \label{eq:partition2}
\end{equation}

The first term in the expansion of the logarithm in Eq.
(\ref{eq:log}) corresponds to the Boltzmann approximation which is
suited only for $m_j>>T$. Such a condition is satisfied for
baryons since their masses are larger than a typical temperature
of the hadron resonance gas which never exceeds  a critical value
$T_c\simeq 200$ MeV required for deconfinement. However, for light
bosons like pions or kaons the quantum statistics is of importance
as the temperature is comparable to their masses. Fig.
(\ref{fig1}--left) shows relative deviations of pion and kaon
multiplicities from their quantum statistics values with
increasing numbers of terms $k$ in the expansion (\ref{eq:log}).
The calculations were performed for $T=157$ and $V=32$ fm$^3$. It
is clear that the  Boltzmann approximation is by far not
sufficient to reproduce the quantum statistics results. The pion
yield under Boltzmann approximation deviates by more than $7\%$
from the  exact quantum statistics result. For kaons this
difference is only $1\%$ due to the larger  mass. For pions, five
terms are needed in the expansion (\ref{eq:log}) to get quantum
statistics value with deviations below $10^{-3}$. For heavier
particles, for instance for protons, the quantum and Boltzmann
statistics differs  by less than $0.1\%$. Thus, in the model
comparison with $e^+e^-$ data we apply quantum statistics for
light mesons and we use Boltzmann statistics for all baryons.

\begin{figure}[htb]
\centering\includegraphics[width=0.54\textwidth]{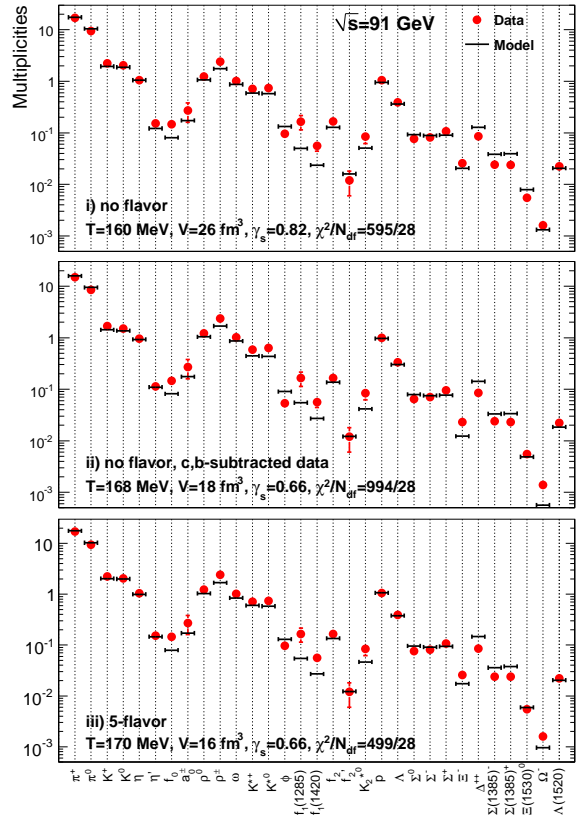}
\caption{Comparison between the best fit thermal model
calculations and experimental hadron multiplicities (sum of
particles and antiparticle yields and for the 2 jets) for $e^+e^-$
collisions at $\sqrt{s}$=91 GeV. The upper panel shows the fit in
an uncorrelated jet scheme  when  including the feed-down
contribution from heavy quarks.  The middle panel is after
subtraction of this contribution.  The results in the lower panel
are obtained in a  correlated  (5-flevour) jet scheme   fit  where
the flavor abundances are extra input parameters from data (see
text). The best thermal model fit parameters are listed for each
case.} \label{fig2}
\end{figure}

\section{Modelling the $e^+e^-$ events within the statistical approach}

The hadron multiplicity calculation within the statistical model
basically proceeds in two steps. First, a primary hadron yield
$N_h^{th}$, is calculated using (\ref{eq:partition1}) and
(\ref{eq:partition2}). A crucial assumption of the model is that
the final yields of all particles are fixed at a common
temperature, the chemical decoupling point. As a second step, all
resonances in the gas which are unstable against strong decays are
allowed to decay into lighter stable hadrons, using appropriate
branching ratios (B) and multiplicities (M) for the decay $j
\rightarrow h$ published by the PDG \cite{Yao:2006px}. The
abundances in the final state are thus determined by
\begin{equation}\label{res}
N_h = N_h^{th} + \sum_j N_j \cdot B(j\rightarrow h) M(j\rightarrow
h)
\end{equation}
where the sum runs over all resonance species.

From Eq. (\ref{res}) it is clear, that the final multiplicity of
stable hadrons depends on the number of resonances used in the
sum. In general one should include  contributions of all known
resonances
 as listed by the  PDG. Fig. (\ref{fig1}-right) shows the
relative change of different particle yields  when applying the
mass cut $M=1.7$ and $M=3.0$ GeV in Eq. (\ref{res}). For mesons
this difference amounts to $5\%$ and is as large as $15\%$ for
baryons. Thus, it is clear that restricting the mass spectrum only
to the resonances with the mass $M<1.7$ GeV might be  not
sufficient at the level af accuracy of data of a few percent  as
is the case in $e^+e^-$ collisions. However, our knowledge of
decay properties of heavier resonances is by far not complete,
which causes systematic uncertainties of the statistical model.
The importance of heavier resonances originating from the Hagedorn
mass spectrum in the analysis of particle production in heavy ion
collisions has been recently analyzed in  \cite{new}. In the
following we account for resonance contributions up to  the mass
$M<3$ GeV.

In general, the resonance gas model formulated in the canonical
ensemble is described by only two basic thermal parameters: the
temperature T and the volume V of the system. To explore a
possible strangeness undersaturation
  we  introduce an additional
parameter $\gamma_s$ into the partition function to account for a
possible deviation of strange particle yields from their chemical
equilibrium values \cite{Becattini:1995if}. When applying the
statistical model to  particle production in $e^+e^-$
annihilations we have to take into account that most hadronic
events  in such collisions  are two-jet events, originating from
quark-antiquark pairs of the five lightest flavors. Since we would
like to address the issue of overall equilibration in these
systems, we  have to specify how the initial quantum numbers are
distributed between  the two-jets. We will consider two scenarios:
an {\sl uncorrelated}  and {\sl correlated} jet scheme.

In an {\sl uncorrelated} jet scheme  each jet is treated as a
fireball with vanishing quantum numbers as fixed by the entrance
channel. It is clear at this point that hadrons from jets with
heavy quarks (c and b) will be greatly underestimated by the model
because of the large Boltzmann suppression factors. In this
approach the issue of equilibration is effectively addressed only
for hadrons with light quarks (u, d, s). It is important to
recognize that the measured yields of these hadrons contain the
contribution from the $e^+e^-$ annihilation events into $c\bar c$
and $b\bar b$. Heavy-quark production is indeed significant and is
very precisely measured, in particular at the $Z_0$ mass, where
the measurements are very well described by the standard model .
Hence, heavy-quark production is manifestly non-thermal in origin.
We therefore consider two cases: i) we fit the data as measured
and  ii) we subtract from the yields of hadrons carrying light
quarks the contribution originating from charm and bottom decays
based on available data for the charmed and bottom hadron
production and their branching ratios.

In  a {\sl correlated} jet scheme (case (iii) in Fig. (\ref{fig2})
) the initial quantum numbers are distributed such that  each jet
carries quantum numbers of either  $S =\pm 1$, $C = \pm 1$, $B
=\pm 1$, or vanishing quantum numbers in the case of $u\bar u$ and
$d\bar d$ jets. The fractions of the quark flavors in hadronic
events \cite{pdg} are external input values, unrelated to the
thermal model (see also Table II in ref. \cite{becattini08}).

\subsection{Model comparison with $e^+e^-$ data at $\sqrt s=91$ GeV}

For the fit procedure we use the complete set of all measured
hadron yields with the exception of those containing charm or
bottom quarks. A $\chi^2$ fit is performed by minimizing
\begin{equation} \chi^2 =
\sum_h\frac{(N_h^{exp}-N_h)^2}{\sigma_h^2}
\end{equation}
as a function of the three parameters T, V and $\gamma_s$, taking
account of
the experimental uncertainties $\sigma_h$.

The resulting best fit to the data for a correlated and
uncorrelated jet scheme  is shown in Fig. (\ref{fig2}). The two
cases of the fit, without and with the subtraction of the
contribution from heavy quarks in an uncorrelated jet scheme  are
also shown in this figure. We first note the overall behavior of
the data, namely an approximately exponential decrease of the
particle yield with an increasing particle mass. Such a behavior
is expected in the hadron resonance gas model due to the Boltzmann
factors, thus indicating the presence of statistical features of
hadron production in elementary collisions. The quantitative
description of the data with the statistical model is, however,
rather poor and certainly no improvement is visible for the case
of subtracting charm and bottom contributions. The low fit quality
is reflected  through the  large $\chi^2$ values per degree of
freedom.
 In addition,   discrepancies between individual  data points and fit
values larger than   5 standard deviations are not rare. There are
also problems in determination of the fit parameters: $T,V$ and
$\gamma_s$. In  $\chi^2$ contour plots, both in $(T,V)$- and
$(T,\gamma_s)$-plane, one notices  strong anticorrelations between
fit parameters \cite{our}.  In addition there is a series of local
minima which makes it difficult to uniquely determine the model
parameters. Such local minima are typical for poor fits and imply
large uncertainties in the determinations of the  fit parameters.

\subsection{Model description of heavy quark hadron production}

In the canonical formulation of the statistical model with  exact
charge\footnote{The charge is consider here to be  any quantum
number related with U(1) symmetry.} conservation the abundances of
charged particles depend crucially on the overall charge  in  a
system. In order to illustrate this let us consider a   model
where only one charge, e.g. charm,  is conserved exactly. In such
a case the multiplicity $<N_i>_{C_i=\pm 1}^C$ of particle $i$ with
mass $m_i$ that carries charge $C_i=\pm 1$ in a system of the
total charge $C$, volume $V$ and temperature $T$  is obtained
under the Boltzmann approximation from \cite{hagedorn}:

\begin{equation}\label{eq:c}
<N_i>_{C_i=\pm 1}^C=V~z_i^{C_i}{{Z_{\mp 1}}\over x}{{I_{C\mp
1}(2Vx)}\over {I_{C}(2Vx)}}
\end{equation}
where
\begin{equation}\label{eq}
z_i^{C_i}={{d_i}\over {2\pi^2}}m_i^2TK_2(m_i/T), ~~~Z_{\pm
1}=\sum_i z_i^{C_i=\pm 1},~~x=\sqrt{Z_1Z_{-1}},
\end{equation}
and where $I_k$ and $K_2$ are Bessel functions and   the argument
of $I_k$   quantifies the total number of charged particle pairs.

From Eq. (\ref{eq:c}) one recognizes an essential difference in
particle yields if the total charge  $C=0$ or $C=\pm 1$. Indeed,
assuming  that  $Vx\leq 1$    and applying an asymptotic expansion
of the Bessel functions one finds e.g. that multiplicities of
particles with  charge $C_i=+1$ is

\begin{equation}\label{eq:1}
<N_i>\simeq V^2z_iZ_{-1},
\end{equation}
if the total charge of a system $C=0$, and
\begin{equation}\label{eq:2}
<N_i>\simeq {{z_i}\over {Z_{+1}}},
\end{equation}
if the total charge of a system $C=+1$.

\begin{figure}[htb]
\centering\includegraphics[width=24.9pc,height=14.4pc]{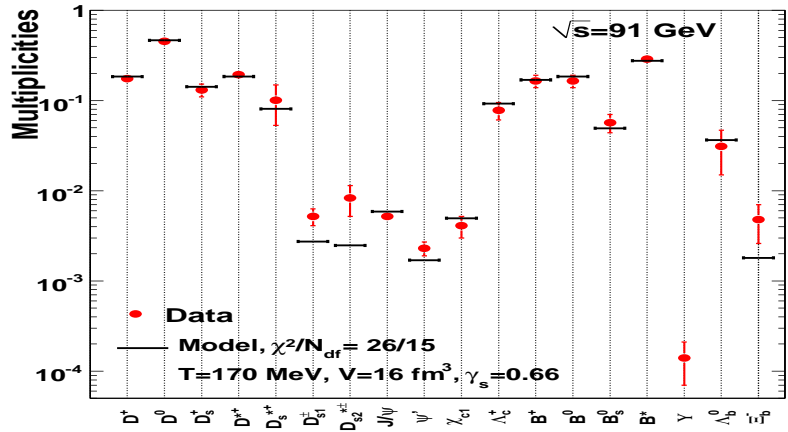}
\caption{Comparison between the  thermal model calculations in a
correlated jet scheme and experimental  data on charmed and
bottomed hadrons for   $e^+e^-$ collisions at $\sqrt{s}$=91 GeV.
Also shown are the model parameters used in the calculations.  }
\label{fig3}\end{figure}

For a charge neutral system the charged particles yield
(\ref{eq:1}) are strongly suppressed due to canonical effects,
since a particle has to be produced in a pair with an antiparticle
in order to fulfill   charge neutrality. This is a well known
"canonical suppression effect" which is crucial e.g. to quantify
"strangeness enhancement" and production in hadron-hadron and
heavy ion collisions at lower energies \cite{review}. For the
total charge $C=\pm 1$ the canonical effect results in enhancement
of particle yields as shown in  Eq. (\ref{eq:2}). The charge
$C=\pm 1$ of a system is here redistributed between all particles
that carry $C_i=\pm 1$ with  weights  given by the ratio of the
thermal phase--space of particle $i$ to the phase--space of all
negatively or positively charged particles. Here, even the
Boltzmann suppression  is to a large extent cancelled out in the
ratio  (see Eq. (\ref{eq:2})).

The above examples  imply  a qualitative difference  in open charm
and bottom production in $e^+e^-$ collisions when using an
uncorrelated and correlated jet scheme scenario. If $C=B=0$ in a
jet then the thermal phase--space of charm and bottom is strongly
suppressed as in Eq. (\ref{eq:1}), leading to large discrepancies
between model results and data. One needs to  check whether in the
correlated jet scheme, where the open charm and bottom are
statistically  enhanced as in Eq. (\ref{eq:1}), the measured
yields are comparable with model predictions.

 In  quantitative
analysis, one calculates the heavy quark yields with the partition
function (\ref{eq:partition1}) rather than with the approximate
Eq. (\ref{eq:2}). Fig. (\ref{fig3})  shows the statistical model
results for charmed and bottom particles  obtained in the
correlated jet scheme. One recognizes  a good description of data
by the canonical statistical model. Thus, if the fireballs related
with each jet are charged then the model description of the
distribution of these charges between different particles  agrees
with data. The value of $\chi^2/N_{df}\simeq 2$ indicates a good
quality of the model description of data. We have to stress that,
in the canonical model, yields of  hidden bottom and charm meson
are still strongly suppressed by the Boltzmann factors, thus their
multiplicities deviate from data. Indeed, in Fig. (\ref{fig3}),
the yield of $Y$ is  larger by several orders of magnitudes than
the model results (not shown in Fig.(\ref{fig3})), indicating a
non-thermal origin of this particle. This is not the case for
$\psi$ since in $e^+e^-$ annihilations at $\sqrt s=91$ GeV as they
are almost entirely originating from decays of bottom
\cite{ourn}.\footnote{ For a more complete discussion on charm and
bottom production see Ref. \cite{ourn}}.

\section{Summary and conclusions}
We have analyzed the experimental data on hadron yields with light
and heavy quarks  in $e^+e^-$ collisions at $\sqrt s= 91$ GeV
within the statistical model. The conservation of five quantum
numbers was included in the framework of the canonical partition
function. Our results qualitatively confirmed the previous finding
\cite{Becattini:1995if} that statistical features are present in
hadron production in $e^+e^-$ annihilation. The resulting
temperature of $160-170$ MeV lies in the bulk expected at the
chemical freezeout in heavy ion collisions at high energy
\cite{aat,review} and agrees with first results obtained in
$e^+e^-$ systems \cite{Becattini:1995if}. Our results on open
charm and bottom hadrons, using the measured $c$ and $b$ fractions
of jets,  showing good agreement with data also confirmed earlier
observation \cite{Becattini:1995if}. However, in view of a rather
poor fit to measured yields of hadrons with light quarks and a
clearly non-thermal origin of the hidden charm and bottom
particles as well as essentially different characteristics of the
collision fireball in $e^+e^-$  and in heavy ion collisions
\cite{our},  the general validity of the thermodynamical approach
to the particle production in $e^+e^-$ annihilation at LEP
energies is not obvious. 

\end{document}